\documentclass[onecolumn,english,american,notitlepage,usenames,dvipsnames,svgnames,table]{revtex4-1}

\usepackage[T1]{fontenc}
\usepackage[latin9]{inputenc}
\usepackage{amsmath}
\usepackage{graphicx}
\usepackage[american]{babel}

\usepackage{esint}
\usepackage{color}
\usepackage{cfr-lm}
\usepackage{microtype}

\let\originalleft\left
\let\originalright\right
\renewcommand{\left}{\mathopen{}\mathclose\bgroup\originalleft}
\renewcommand{\right}{\aftergroup\egroup\originalright}

\makeatletter

\begin{document}

\title{Simulation of Scalar Field Theories with Complex Actions}

\author{Leandro Medina and Michael C.~Ogilvie}

\address{Dept.~of Physics, Washington University, St.~Louis, MO 63130 USA}
\email{mco@wustl.edu, leandro@wustl.edu}

\begin{abstract}
We develop a method for the simulation of scalar field theories with
complex actions which is local, simple to implement and can be used
in any number of space-time dimensions. For model systems satisfying
the $\mathcal{PT}$ symmetry condition $L^{*}(\phi)=L(-\phi)$, the
complex weight problem is reduced to a sign problem. The sign problem
is eliminated completely for a large subclass of these models; this
class includes models within the $i\phi^{3}$ universality class, and
also models with nonzero chemical potential. Simulations of models
from this subclass show a rich set of behaviors. Propagators may
exhibit damped oscillations, indicating a clear violation of spectral
positivity. Modulated phases occur in some models, exhibiting striping
and other pattern-forming behaviors. These field theory models are
connected to complex systems where pattern formation occurs because
of competition between interactions at two different length scales.
\end{abstract}
\maketitle

The sign problem is a significant obstacle in theoretical physics.
Computational methods for averaging over positive weights have
been highly developed in lattice field theory and statistical physics, 
but averaging with real or complex weights
has proven difficult. Two examples of sign problems
in lattice field theory are associated with QCD with nonzero chemical potential
$\mu$~\cite{deForcrand:2010ys,Aarts:2015tyj}, and the $i\phi^{3}$ field
theory, which determines the critical indices of the Lee-Yang edge
transition~\cite{Fisher:1978pf}.
The sign problem in QCD with $\mu\ne0$ is particularly vexing because
many interesting questions in particle physics, nuclear physics and
astrophysics depend on the properties of hadronic matter at nonzero
baryon density. Lattice simulations have been enormously successful
in elucidating the phase structure of QCD and related theories at
finite temperature $T$, but a complete first-principles understanding
of the QCD phase diagram in the $\mu-T$ plane remains unrealized.
The difficult problem of simulating QCD at nonzero $\mu$ was clear
decades ago~\cite{Hands:1999md}, but a satisfactory algorithm has
remained elusive. There are several approaches to the sign problem
of finite density QCD under active development, including methods
based on duality~\cite{Gattringer:2016kco}, Lefschetz thimbles~\cite{Cristoforetti:2012su},
and complex Langevin equations~\cite{Seiler:2017wvd}. 

A large class of lattice field theories with sign problems satisfy
a discrete symmetry of the form $L^{*}(\phi)=L(-\phi)$, which
is a form of $\mathcal{PT}$ symmetry~\cite{Bender:1998ke,Bernard:2001wh,Bender:2007nj}. 
We will present in this letter a procedure for finding dual forms
of the partition function $Z$ for a large class of such $\mathcal{PT}$-symmetric
scalar field theories; a duality transformation yields representations
with real local actions which are easily simulated by standard lattice
field theory methods in any number of dimensions. This class includes models
in the $i\phi^{3}$ universality class as well as charged scalar fields
with nonzero chemical potential. Models in this class exhibit a rich
set of possible behaviors. Because their transfer matrices are non-Hermitian,
they may have complex eigenvalues. This leads to damped
oscillations of correlation functions, a behavior well-known
in the context of improved actions~\cite{Luscher:1984is}. This
represents a loss of spectral positivity, but is seen in many physical
systems, \emph{e.g.}~liquids. In some cases, spatially modulated phases
occur~\cite{Meisinger:2012va}.

Our starting point is a model with a single scalar lattice field $\chi$
and an Euclidean action of the form 
\begin{equation}
S(\chi)=\sum_{x}\left[\tfrac{1}{2}(\partial_{\mu}\chi(x))^{2}+V(\chi(x))\right]
\end{equation}
where the sum is over all lattice sites $x$ and $\partial_{\mu}\chi(x)=\chi(x+\hat{\mu})-\chi(x)$.
We take the potential $V$ to satisfy $V(-\chi)=V(\chi)^{*}$. Because
of this condition, the Fourier transform $\tilde{w}(\tilde{\chi})$
of $w(\chi)\equiv\exp[-V(\chi)]$ with respect to $\chi$
is real. If $\tilde{w}(\tilde{\chi})$ is everywhere positive, we
say that the \emph{dual positivity condition} is satisfied and define
a real function $\tilde{V}(\tilde{\chi})=-\log(\tilde{w}(\tilde{\chi}))$.
The partition function is
\begin{equation}
Z[h]=\int\prod_{x}d\chi(x)\exp\left[-S(\chi(x))+\sum_{x}ih(x)\chi(x)\right]
\end{equation}
where $h(x)$ is an arbitrary source. We rewrite $Z$ as
\begin{equation}
Z=\int\prod_{x}d\pi_{\mu}(x)d\tilde{\chi}(x)d\chi(x)\exp\left\{ -\sum_{x}\left[\tfrac{1}{2}\pi_{\mu}^{2}(x)+i\pi_{\mu}(x)\partial_{\mu}\chi(x)+\tilde{V}(\tilde{\chi}(x))+i\chi(\tilde{\chi}(x)+h(x))\right]\right\} .
\end{equation}
After a lattice integration by parts, the integral over $\phi$ yields
\begin{equation}
Z=\int\prod_{x}d\pi_{\mu}(x)\exp\left\{ -\sum_{x}\left[\tfrac{1}{2}\pi_{\mu}^{2}(x)+\tilde{V}(\partial\cdot\pi(x)-h(x))\right]\right\} .
\end{equation}
This represents a dual form of the partition function for fields defined
on the real line, similar to dual forms for models with fields defined
on compact manifolds. In the more general case where the dual positivity
condition does not hold, the same procedure reduces the problem of
simulating a complex action to the simulation of a system with a true
sign problem. The strategy for simulation of the model is now clear:
one simulates the new action
\begin{equation}
\tilde{S}[\pi_{\mu}]=\sum_{x}\left[\tfrac{1}{2}\pi_{\mu}^{2}(x)+\tilde{V}(\partial\cdot\pi(x))\right].
\end{equation}
There is a clear extension to the case of more than one $\mathcal{PT}$-symmetric
scalar field. Correlation functions of the original field $\chi$
may be obtained in the usual way by differentiation with respect to
$h(x)$. In particular, non-coincident correlation functions of $\chi$
are obtained in the new representation using the field $i\tilde{V}'(\partial\cdot\pi(x))$
. Note that the expectation value of $\chi$ is either zero or purely
imaginary.

\begin{figure}[t]
\begin{centering}
\includegraphics{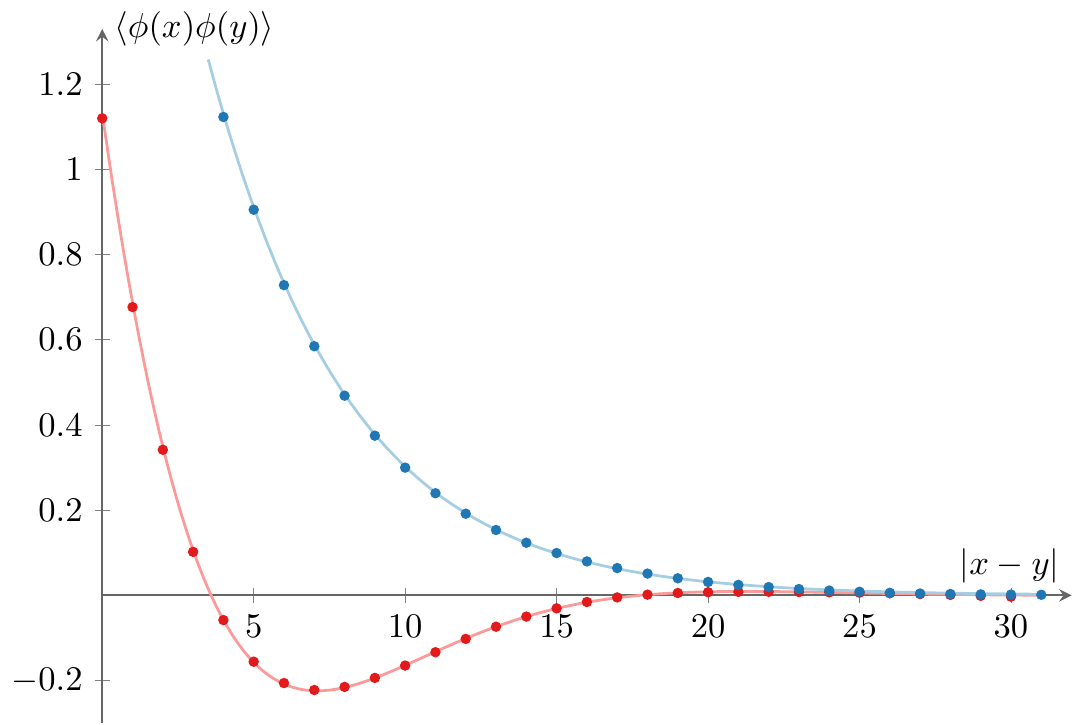}
\par\end{centering}
\caption{\label{fig:icqprop}Propagator $\langle\phi(x)\phi(y)\rangle$ as
a function of $|x-y|$ for the $d=1$ ICQ model on a lattice
with $256$ sites. In both curves, $m_{\phi}^{2}=0.001$ and $g=0.1$.
The upper (blue) curve corresponds to $m_{\chi}^{2}=0.250$, while
the lower (red) curve has $m_{\chi}^{2}=0.002$. The lines represent
the analytical form of the continuum result; the errors bars on all
points are smaller than the points themselves.}

\end{figure}

The allowed forms of $V$ are constrained by Bochner's
theorem, which states that a function $\tilde{w}(k)$ on a locally
compact Abelian group is positive if and only if $w(x)$ is positive-definite,
that is, the matrix $w(x_{j}-x_{k})$ has positive eigenvalues for
any choice of the set $\{x_{j}\}$. This theorem is easily understood
from the formula 
\begin{equation}
\int dx\,\psi^{*}(x)w(x)\psi(x)=\int\frac{dk}{2\pi}\frac{dq}{2\pi}\tilde{\psi}^{*}(k)\tilde{w}(k-q)\tilde{\psi}(q).
\end{equation}
The simplest of the constraints imposed by Bochner's theorem is $V(x)+V(-x)>2V(0)$.
This constraint excludes the double-well potential and other potentials
that lead to conventional spontaneous symmetry breaking at tree level.
This is consistent with $\mathcal{PT}$ symmetry which requires $\langle\chi\rangle^{*}=-\langle\chi\rangle$,
\emph{i.e.}, that $\langle\phi\rangle$ be purely imaginary. We do
not view this limitation as fundamental, because this constraint does
not appear in other duality-based treatments of complex actions~\cite{Gattringer:2012df}.

It is simplest to take the form of the dual potential $\tilde{V}$
as given, and determine the parameters of $V$ from it. If one parametrizes
$V$ as a polynomial in $\phi$, $V=\sum_{n}g_{n}\phi^{n}/n!$ , then
the coefficients $g_{n}$ are naturally obtained from the generating
functional of the zero-dimensional dual theory defined by
\begin{equation}
g_{n}\equiv\left.\frac{\partial^{n}V}{\partial\chi^{n}}\right|_{\chi=0}=-\left(i\right)^{n}\left\langle \tilde{\chi}^{n}\right\rangle _{c},
\end{equation}
the cumulant of the dual variable $\tilde{\chi}$ averaged with weight
$\tilde{w}(\tilde{\chi})$. The mass parameter of $V$ is given by
$g_{2}=\langle\tilde{\chi}^{2}\rangle_{c}=\langle\tilde{\chi}^{2}\rangle-\left\langle \tilde{\chi}\right\rangle ^{2}$
and is therefore always positive. It follows that in the dual theory
the case $g_{2}=0$ can only be obtained by a limiting process, and
the region with $g_{2}<0$ is not directly accessible. Note that $g_{2}$
is the bare parameter, defined at the scale of a lattice spacing.
For a generic dual potential $\tilde{V},$ the coupling $g_{3}$ is
nonzero and imaginary, so critical behavior is naturally in the $i\phi^{3}$
universality class. 

It is very interesting to consider the coupling of $\mathcal{PT}$-symmetric
scalar fields to normal fields~\cite{Bender:2007wb}. We write the
action as
\begin{equation}
S\left(\phi\right)=\sum_{x}\left[\tfrac{1}{2}(\partial_{\mu}\phi(x))^{2}+\tfrac{1}{2}(\partial_{\mu}\chi(x))^{2}+V(\phi(x),\chi(x))\right]
\end{equation}
where the potential obeys the condition $V(\phi(x),\chi(x))^{*}=V(\phi(x),-\chi(x))$.
Following similar steps applied to $\chi$ as those given above, we
arrive at a dual action of the form
\begin{equation}
\tilde{S}=\sum_{x}\left[\tfrac{1}{2}(\partial_{\mu}\phi(x))^{2}+\tfrac{1}{2}\pi_{\mu}^{2}(x)+\tilde{V}(\phi(x),\partial\cdot\pi(x)-h(x))\right].
\end{equation}
As in the case of a single field, expectation values involving $\chi$
can be obtained from $i\partial\tilde{V}'(\phi,\partial\cdot\pi)/\partial(\partial\cdot\pi)$.
The case of a complex field with a nonzero chemical potential $\mu$
is similar, but the complex contribution to action arises in the kinetic
term rather than in $V$, and requires a separate treatment.

In this class of models $\phi$ and $\chi$ play roles similar to
the real and imaginary parts of the Polyakov loop, $P_{R}$ and $P_{I}$,
in QCD at finite temperature $T$ and chemical potential $\mu$. The
Polyakov loop $P=P_{R}+iP_{I}$ is associated with the free energy
required to insert a very heavy quark into the system via $\langle P\rangle=\exp(-F_{Q}/T)$;
$\langle P^{*}\rangle$ is related to the free energy $F_{\bar{Q}}$
for insertion of a heavy antiquark. When $\mu=0$, $P_{R}$ develops
a real expected value and $\langle P_{I}\rangle=0$ such that $F_{Q}=F_{\bar{Q}}$.
When $\mu\ne0$, QCD has a sign problem, and a variety of techniques
show that $P_{I}$ acquires an imaginary expectation value~\cite{DeGrand:1983fk,Hands:2010zp,Nishimura:2014kla,Nishimura:2014rxa,Akerlund:2016myr}. This implies
that $\langle P\rangle\ne\langle P^{*}\rangle$ and $F_{Q}\ne F_{\bar{Q}}$.
Both phenomenological models~\cite{Nishimura:2014rxa,Nishimura:2014kla}
as well as simplified lattice models of QCD at nonzero density~\cite{Akerlund:2016myr}
show that correlation functions may exhibit damped oscillatory behavior
for some range of parameters.
\begin{figure}[t]
\begin{centering}
\includegraphics{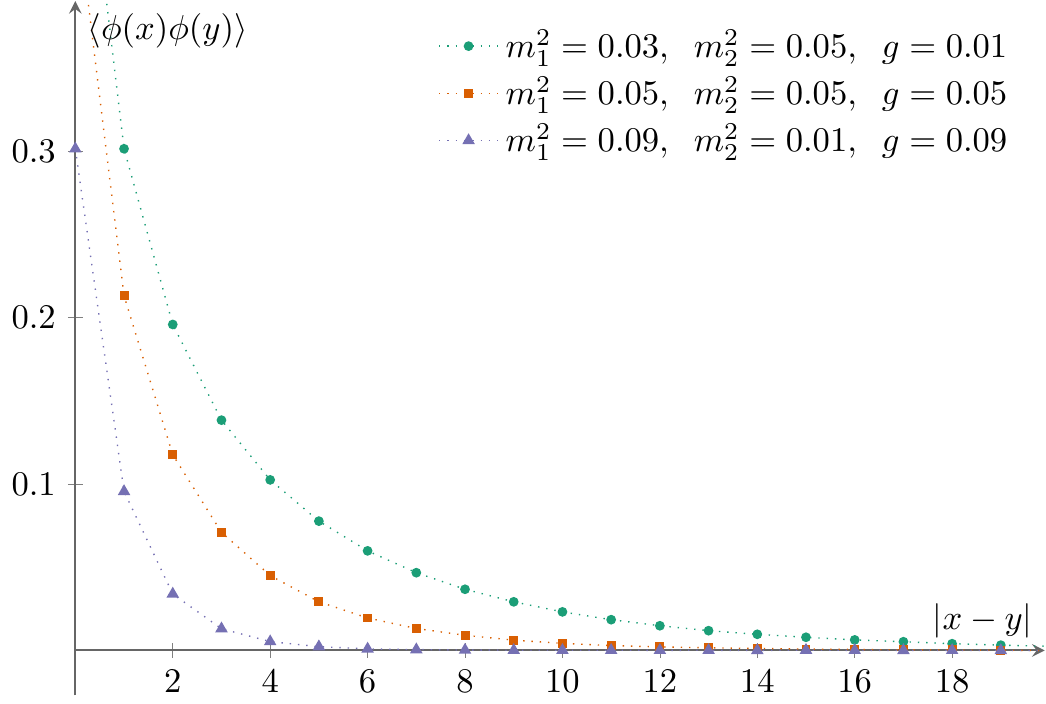}
\par\end{centering}
\caption{\label{fig:icyprop}Typical results for the propagator $\langle\phi(x)\phi(y)\rangle$
as a function of $|x-y|$ for the $d=2$ ICY model for three different
parameter sets on a $64^{2}$ lattice.}
\end{figure}

As a demonstration of the technique and the variety of results which
may be obtained, we consider three different models. In all three
models, $S$ is quadratic in $\chi$ so that both $V$ and $\tilde{V}$
are known analytically. Our first model is exactly solvable but displays
nontrivial behavior. This imaginary-coupled quadratic (ICQ) model
has a potential $V$ of the form $V(\phi,\chi)=m_{\phi}^{2}\phi^{2}/2+m_{\chi}^{2}\chi^{2}/2-ig\phi\chi$.
The eigenvalues of the mass matrix are given by ${\big(m_{\phi}^{2}+m_{\chi}^{2}\pm\sqrt{\vphantom{m^\phi_o}\smash{(m_{\phi}^{2}-m_{\chi}^{2})^{2}_{\vphantom{\chi}}-4g^{2}}}\big)}/2$,
so there are either two real masses or a complex conjugate pair, as
required by the $\mathcal{PT}$ symmetry of the model. A quantum mechanical
model of this form was considered in~\cite{Bender:2007wb,Bender:2016vdo}.
The dual potential takes the form $\tilde{V}(\phi,\partial\cdot\pi)=m_{\phi}^{2}\phi^{2}/2+(\partial\cdot\pi-g\phi)^{2}/2m_{\chi}^{2}$.
In figure~\ref{fig:icqprop}, we show results for one-dimensional
simulations of the ICQ model in the two different regions on a lattice
of size $N=256$: the difference in behavior is striking between the
upper curve where there are two real masses, and the lower curve where
there is a complex conjugate mass pair. Similar results were obtained
in two-dimensional simulations. The lines represent the analytical
form of the continuum result for the propagators, and the error bars
on the points are smaller than the points themselves. The nonmonotonicity
of the lower curve makes the violation of spectral positivity obvious;
in fact the lower curve is a damped sinusoid. The analytical result
for the upper curve shows that it is the \emph{difference} of two decaying
exponentials, and therefore also violates spectral positivity. It
is interesting to note that with the definition $\psi=(\partial\cdot\pi-g\phi)/m_{2}^{2}$,
the equations of motion obtained from $\tilde{S}$ can be reduced
to a set of real linear equations for $\phi$ and $\psi$. These equations
may be derived from a Lagrangian of the form
\begin{equation}
\tfrac{1}{2}(\partial\phi)^{2}+\tfrac{1}{2}m_{\phi}^{2}\phi^{2}-\tfrac{1}{2}(\partial\psi)^{2}-\tfrac{1}{2}m_{\chi}^{2}\psi^{2}-g\phi\psi
\end{equation}
but this Lagrangian is not suitable for lattice simulation due to
the negative quadratic terms.

A second interesting model is the imaginary coupling Yukawa (ICY)
model, where the potential has the form $V(\phi,\chi)=m_{\phi}^{2}\phi^{2}/2+m_{\chi}^{2}\chi^{2}/2-ig\chi\phi^{2}$.
This in turn leads to a dual potential $\tilde{V}(\phi,\partial\cdot\pi)=m_{\phi}^{2}\phi^{2}/2+(\partial\cdot\pi-g\phi^{2})^{2}/2m_{\chi}^{2}$.
We show in figure~\ref{fig:icyprop}, the typical behavior of the
propagator $\langle\phi(x)\phi(y)\rangle$ as a function of $|x-y|$
for the $d=2$ ICY model. An extensive search indicated no signs for
a region of parameter space with complex conjugate mass pairs, 
but we were not able to rule out violations of spectral positivity of
the type seen in the ICQ model when both masses are real. It seems
possible that this is a model where masses are always real. This is
consistent with the large-$m_{\chi}$ limit: After the rescaling $\pi_{\mu}\rightarrow m_{\chi}\pi_{\mu}$
and the definition $\lambda=g^{2}/m_{\chi}^{2}$ we can take the limit
$m_{\chi}\rightarrow\infty$ to obtain a potential $\lambda\phi^{4}/2$.
This is a bosonic form of a familiar argument for fermions: the $ig\chi\phi^{2}$
interaction is repulsive and in the large $m_{\chi}$ limit becomes
a repulsive four-boson interaction~\cite{Gross:1974jv}.

Our third example also generalizes the first: we take $V(\phi,\chi)=U(\phi)+m_{\chi}^{2}\chi^{2}/2-ig\chi\phi$,
leading to $\tilde{V}(\phi,\partial\cdot\pi)=U(\phi)+(\partial\cdot\pi-g\phi)^{2}/2m_{\chi}^{2}$.
The potential $U$ can be chosen to give a first-order or second-order
transition as a function of its parameters when $g=0$. We will consider
here the specific case of the imaginary-coupled double well (ICDW)
model, where the potential has the form $U(\phi)=\lambda(\phi^{2}-v^{2})^{2}$.
Because the field $\chi$ enters quadratically, it may be integrated
out, yielding an effective action of the form
\begin{equation}
S=\sum_{x}\left[\tfrac{1}{2}(\partial_{\mu}\phi(x))^{2}+U(\phi)\right]+\frac{g^{2}}{2}\sum_{x,y}\phi(x)\Delta(x-y)\phi(x)
\end{equation}
where $\Delta(x)$ is a free Euclidean propagator with mass $m_{\chi}$,
$i.e.$, a Yukawa potential. This additional term in the action acts
to suppress spontaneous symmetry breaking. Models of this type have
been used to model a wide variety of physical systems and are known
to produce spatially modulated phases~\cite{Seul476,Nussinov:1999fu,PhysRevE.66.066108,PhysRevLett.100.246402}.
In this class of models the complex form of
the action intermediates between a real local form and a real quasilocal
form.

\begin{figure}[t]
\begin{centering}
\begin{minipage}[t]{0.21\columnwidth}%
\begin{center}
\includegraphics[width=0.99\columnwidth]{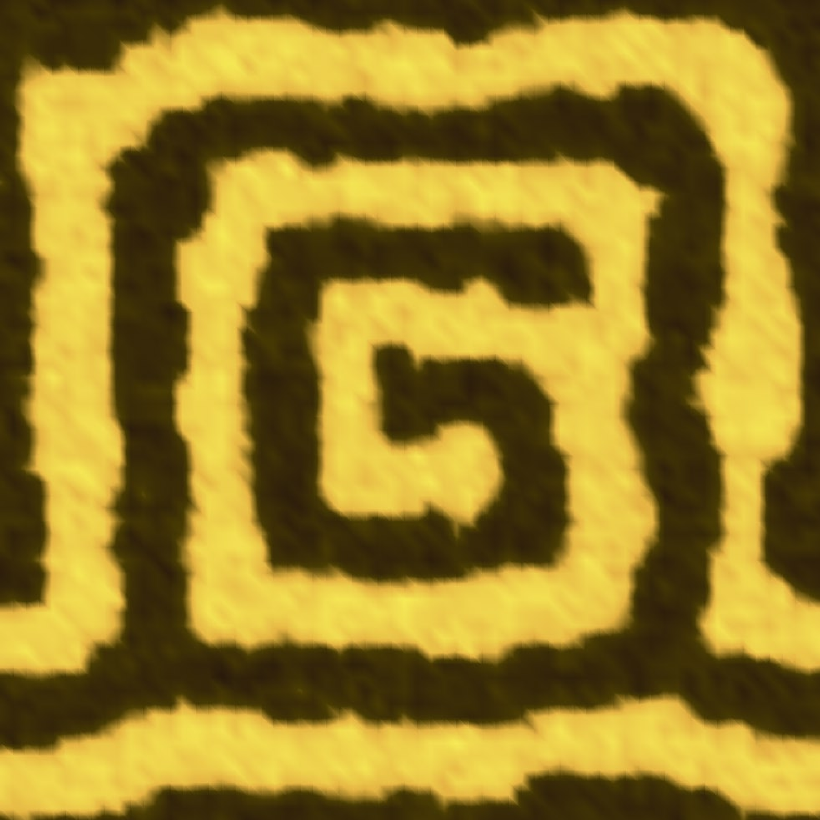}
\par\end{center}%
\end{minipage}%
\begin{minipage}[t]{0.21\columnwidth}%
\begin{center}
\includegraphics[width=0.99\columnwidth]{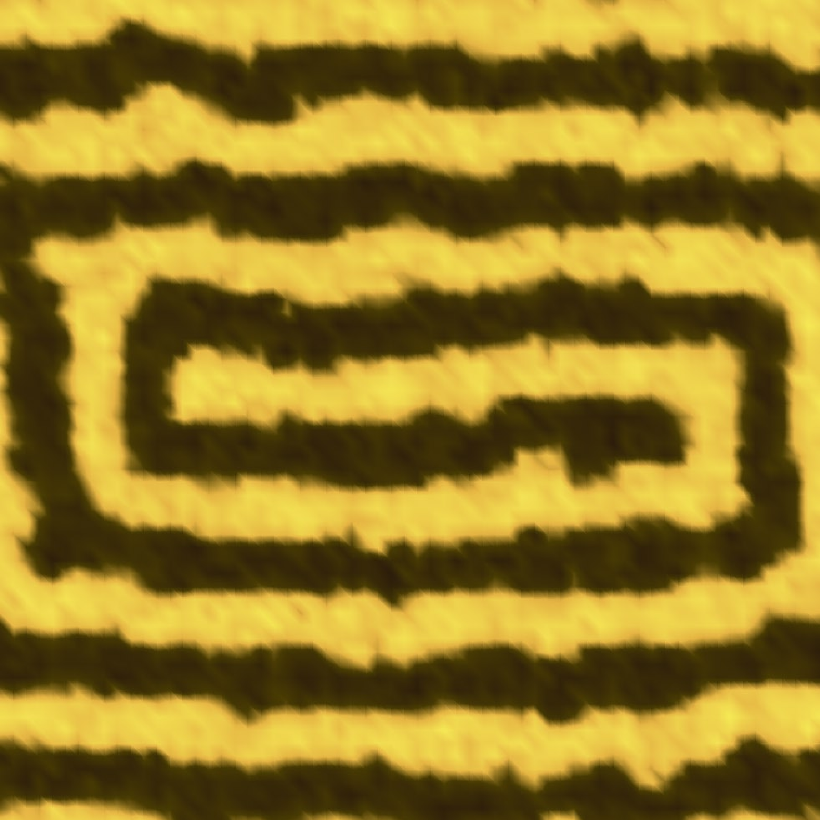}
\par\end{center}%
\end{minipage}%
\begin{minipage}[t]{0.21\columnwidth}%
\begin{center}
\includegraphics[width=0.99\columnwidth]{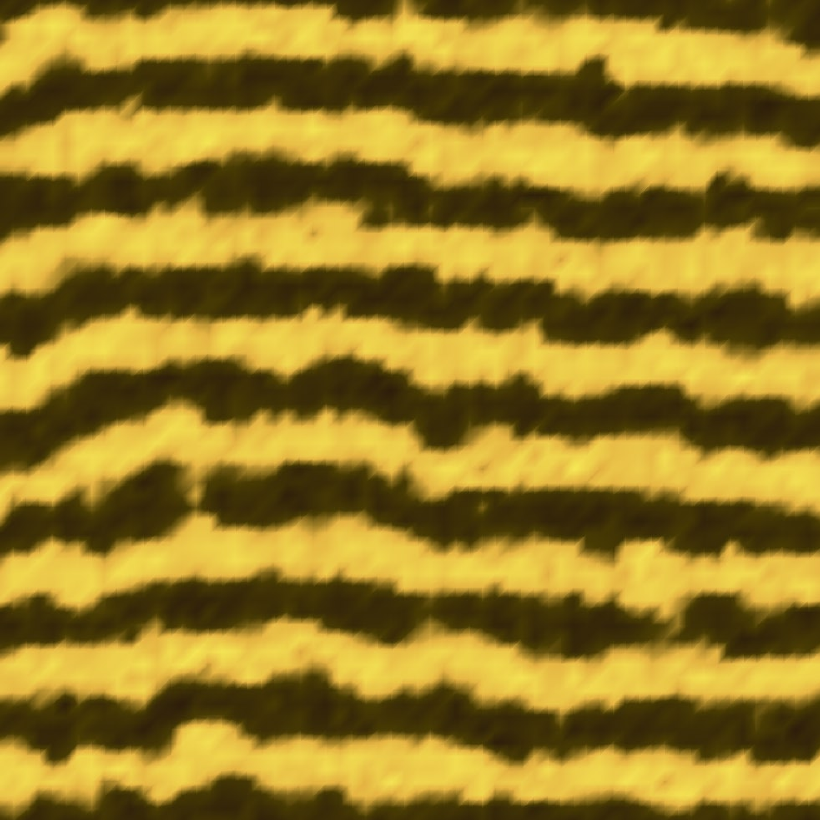}
\par\end{center}%
\end{minipage}
\par\end{centering}
\begin{centering}
%\vspace{0.2mm}
%\vspace{-0.4mm}
\begin{minipage}[t]{0.21\columnwidth}%
\begin{center}
\includegraphics[width=0.99\columnwidth]{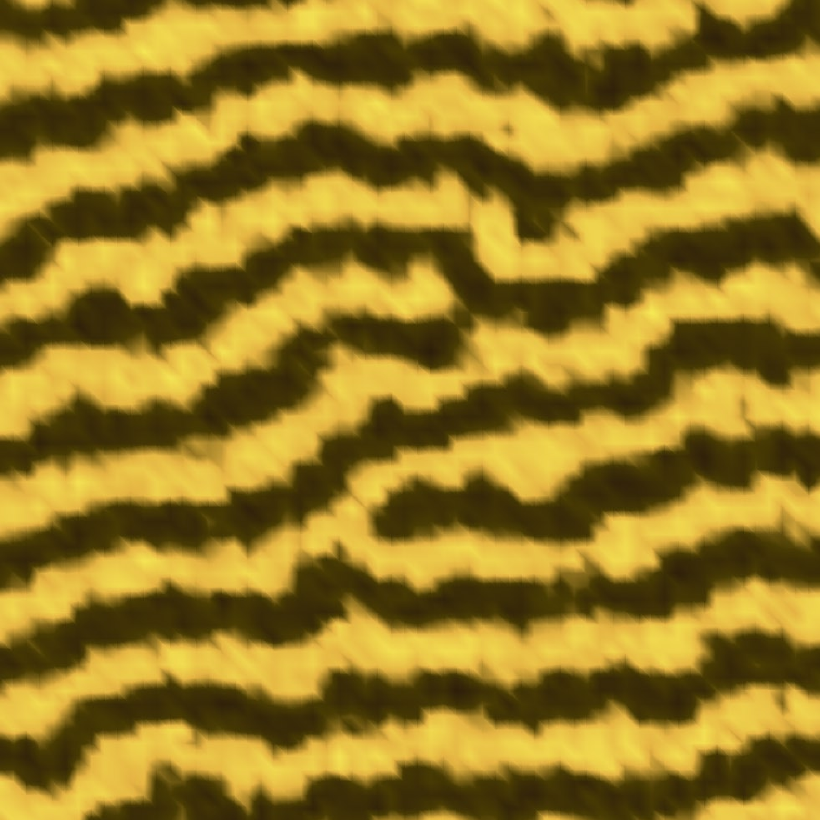}
\par\end{center}%
\end{minipage}%
\begin{minipage}[t]{0.21\columnwidth}%
\begin{center}
\includegraphics[width=0.99\columnwidth]{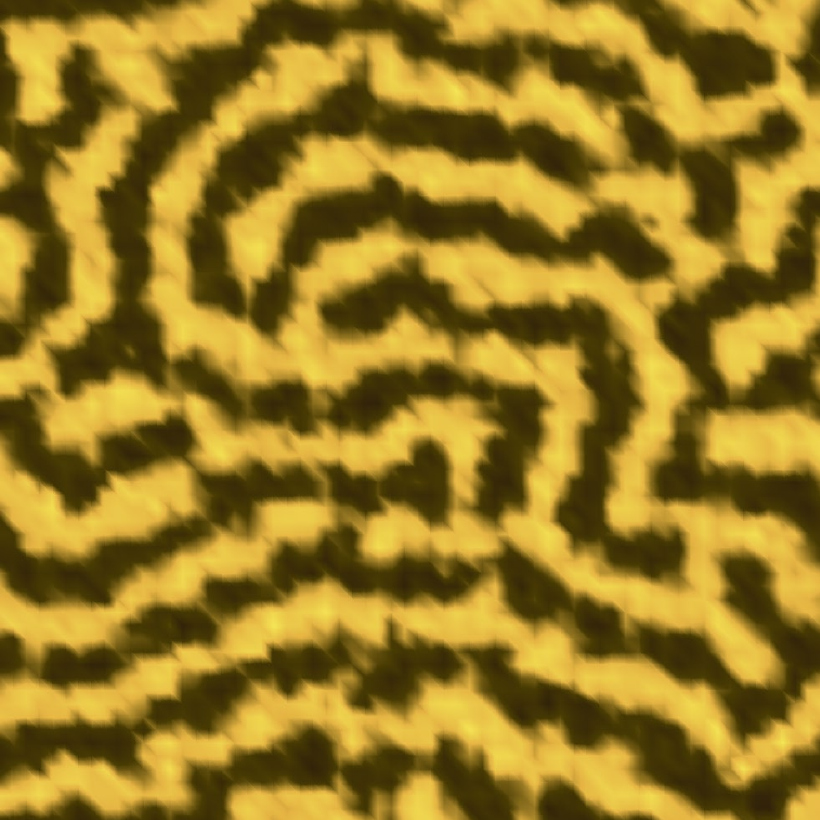}
\par\end{center}%
\end{minipage}%
\begin{minipage}[t]{0.21\columnwidth}%
\begin{center}
\includegraphics[width=0.99\columnwidth]{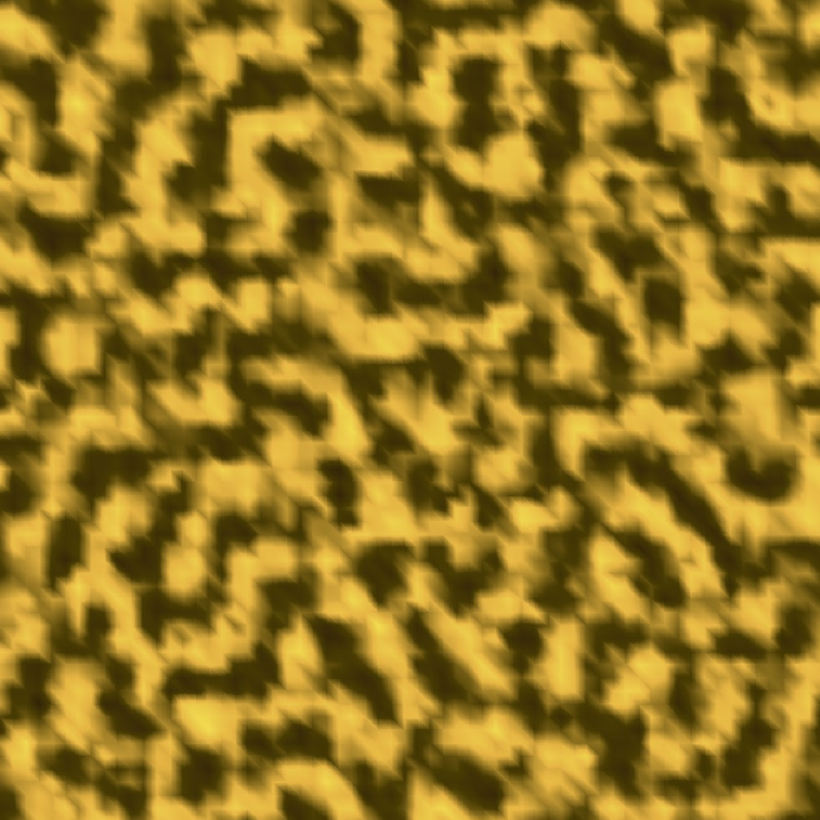}
\par\end{center}%
\end{minipage}
\par\end{centering}
\caption{\label{fig:Configuration-snapshots}Configuration snapshots of $\phi$
in the $d=2$ ICDW model on a $64^{2}$ lattice for several values
of $g$. From left to right, top to bottom: $g=0.9,\,1.0,\,1.1,\,1.2,\,1.3$
and $1.5$. The other parameters are $m_{\chi}^{2}=0.5$, $\lambda=0.1$
and $v=3$.}
\end{figure}

Constant solutions which extremize $S$ can be found by minimizing
$U(\phi)+g^{2}\phi^{2}/2m_{\chi}^{2}$. Linearizing the
equation of motion around such a solution $\phi_{0}$, the inverse
propagator is found to be
\begin{equation}
p^{2}+U''\left(\phi_{0}\right)+\frac{g^{2}}{p^{2}+m_{\chi}^{2}}
\end{equation}
which has a minimum away from zero when $g>m_{\chi}^{2}$, given by
$p_{min}^{2}=g-m_{\chi}^{2}$. For this value of $p^{2}$, the inverse
propagator has the value $2g-m_{\chi}^{2}+U''(\phi_{0})$.
As long as this quantity is positive, the constant solution is stable
to fluctuations at $p^{2}=p_{min}^{2}$. On the other hand, if the
two conditions
\begin{align}
U''\left(\phi_{0}\right)+\frac{g^{2}}{m_{\chi}^{2}} & >  0\\
2g-m_{\chi}^{2}+U''\left(\phi_{0}\right) & <  0
\end{align}
are simultaneously satisfied, then $\phi_{0}$ will be unstable to
modulated behavior with wavenumber $p_{min}$. For $U(\phi)$
a double well, this instability leads to a region where spatially
modulated behavior occurs in lattice simulations of the dual form
of the theory, as shown in the configuration snapshots for $d=2$
in figure~\ref{fig:Configuration-snapshots}. Light and dark portions
represent positive and negative values of $\phi$, respectively. The
length scale of the modulations decreases as $g$ increases and the
system moves toward restoration of the broken symmetry. Similar behavior
is seen for $d=1$ and $d=3$.

For scalar field theories with sign problems satisfying the dual positivity
condition, the methods developed here enable straightforward simulation
with a real local action. Relatively simple models show complicated
behaviors that do not occur in conventional field theories, such as
complex conjugate mass eigenstates and spatially modulated phases.
These models and our simulation method provide a benchmark against
which other simulation methods and analytical techniques can be tested.
This method allows for simulation of models in the $i\phi^{3}$ universality
class as well as field theories with a nonzero chemical potential;
we plan to return to both these topics in subsequent publications.

\bibliographystyle{unsrtnat}
\bibliography{simqft}

\begin{thebibliography}{25}
\providecommand{\natexlab}[1]{#1}
\providecommand{\url}[1]{\texttt{#1}}
\expandafter\ifx\csname urlstyle\endcsname\relax
  \providecommand{\doi}[1]{doi: #1}\else
  \providecommand{\doi}{doi: \begingroup \urlstyle{rm}\Url}\fi

\bibitem[de~Forcrand(2009)]{deForcrand:2010ys}
Philippe de~Forcrand.
\newblock {Simulating QCD at finite density}.
\newblock \emph{PoS}, LAT2009:\penalty0 010, 2009.

\bibitem[Aarts(2016)]{Aarts:2015tyj}
Gert Aarts.
\newblock {Introductory lectures on lattice QCD at nonzero baryon number}.
\newblock \emph{J. Phys. Conf. Ser.}, 706\penalty0 (2):\penalty0 022004, 2016.
\newblock \doi{10.1088/1742-6596/706/2/022004}.

\bibitem[Fisher(1978)]{Fisher:1978pf}
M.~E. Fisher.
\newblock {Yang-Lee Edge Singularity and $\phi^3$ Field Theory}.
\newblock \emph{Phys. Rev. Lett.}, 40:\penalty0 1610--1613, 1978.
\newblock \doi{10.1103/PhysRevLett.40.1610}.

\bibitem[Hands et~al.(1999)Hands, Kogut, Lombardo, and Morrison]{Hands:1999md}
Simon Hands, John~B. Kogut, Maria-Paola Lombardo, and Susan~E. Morrison.
\newblock {Symmetries and spectrum of SU(2) lattice gauge theory at finite
  chemical potential}.
\newblock \emph{Nucl. Phys.}, B558:\penalty0 327--346, 1999.
\newblock \doi{10.1016/S0550-3213(99)00364-8}.

\bibitem[Gattringer and Langfeld(2016)]{Gattringer:2016kco}
Christof Gattringer and Kurt Langfeld.
\newblock {Approaches to the sign problem in lattice field theory}.
\newblock \emph{Int. J. Mod. Phys.}, A31\penalty0 (22):\penalty0 1643007, 2016.
\newblock \doi{10.1142/S0217751X16430077}.

\bibitem[Cristoforetti et~al.(2012)Cristoforetti, Di~Renzo, and
  Scorzato]{Cristoforetti:2012su}
Marco Cristoforetti, Francesco Di~Renzo, and Luigi Scorzato.
\newblock {New approach to the sign problem in quantum field theories: High
  density QCD on a Lefschetz thimble}.
\newblock \emph{Phys. Rev.}, D86:\penalty0 074506, 2012.
\newblock \doi{10.1103/PhysRevD.86.074506}.

\bibitem[Seiler(2017)]{Seiler:2017wvd}
Erhard Seiler.
\newblock {Status of Complex Langevin}.
\newblock In \emph{{35th International Symposium on Lattice Field Theory
  (Lattice 2017) Granada, Spain, June 18-24, 2017}}, 2017.
\newblock URL
  \url{http://inspirehep.net/record/1620215/files/arXiv:1708.08254.pdf}.

\bibitem[Bender and Boettcher(1998)]{Bender:1998ke}
Carl~M. Bender and Stefan Boettcher.
\newblock {Real spectra in nonHermitian Hamiltonians having PT symmetry}.
\newblock \emph{Phys. Rev. Lett.}, 80:\penalty0 5243--5246, 1998.
\newblock \doi{10.1103/PhysRevLett.80.5243}.

\bibitem[Bernard and Savage(2001)]{Bernard:2001wh}
Claude~W. Bernard and Van~M. Savage.
\newblock {Numerical simulations of PT symmetric quantum field theories}.
\newblock \emph{Phys. Rev.}, D64:\penalty0 085010, 2001.
\newblock \doi{10.1103/PhysRevD.64.085010}.

\bibitem[Bender(2007)]{Bender:2007nj}
Carl~M. Bender.
\newblock {Making sense of non-Hermitian Hamiltonians}.
\newblock \emph{Rept. Prog. Phys.}, 70:\penalty0 947, 2007.
\newblock \doi{10.1088/0034-4885/70/6/R03}.

\bibitem[Luscher and Weisz(1984)]{Luscher:1984is}
M.~Luscher and P.~Weisz.
\newblock {Definition and General Properties of the Transfer Matrix in
  Continuum Limit Improved Lattice Gauge Theories}.
\newblock \emph{Nucl. Phys.}, B240:\penalty0 349--361, 1984.
\newblock \doi{10.1016/0550-3213(84)90270-0}.

\bibitem[Meisinger and Ogilvie(2013)]{Meisinger:2012va}
Peter~N. Meisinger and Michael~C. Ogilvie.
\newblock {PT Symmetry in Classical and Quantum Statistical Mechanics}.
\newblock \emph{Phil. Trans. Roy. Soc. Lond.}, A371:\penalty0 20120058, 2013.
\newblock \doi{10.1098/rsta.2012.0058}.

\bibitem[Gattringer and Kloiber(2013)]{Gattringer:2012df}
Christof Gattringer and Thomas Kloiber.
\newblock {Lattice study of the Silver Blaze phenomenon for a charged scalar
  $\phi^4$ field}.
\newblock \emph{Nucl. Phys.}, B869:\penalty0 56--73, 2013.
\newblock \doi{10.1016/j.nuclphysb.2012.12.005}.

\bibitem[Bender and Jones(2008)]{Bender:2007wb}
Carl~M. Bender and Hugh~F. Jones.
\newblock {Interactions of Hermitian and non-Hermitian Hamiltonians}.
\newblock \emph{J. Phys.}, A41:\penalty0 244006, 2008.
\newblock \doi{10.1088/1751-8113/41/24/244006}.

\bibitem[DeGrand and DeTar(1983)]{DeGrand:1983fk}
Thomas~A. DeGrand and Carleton~E. DeTar.
\newblock {Phase Structure of {QCD} at High Temperature With Massive Quarks and
  Finite Quark Density: A $Z$(3) Paradigm}.
\newblock \emph{Nucl. Phys.}, B225:\penalty0 590, 1983.
\newblock \doi{10.1016/0550-3213(83)90536-9}.

\bibitem[Hands et~al.(2010)Hands, Hollowood, and Myers]{Hands:2010zp}
Simon Hands, Timothy~J. Hollowood, and Joyce~C. Myers.
\newblock {QCD with Chemical Potential in a Small Hyperspherical Box}.
\newblock \emph{JHEP}, 07:\penalty0 086, 2010.
\newblock \doi{10.1007/JHEP07(2010)086}.

\bibitem[Nishimura et~al.(2015)Nishimura, Ogilvie, and
  Pangeni]{Nishimura:2014kla}
Hiromichi Nishimura, Michael~C. Ogilvie, and Kamal Pangeni.
\newblock {Complex Saddle Points and Disorder Lines in QCD at finite
  temperature and density}.
\newblock \emph{Phys. Rev.}, D91\penalty0 (5):\penalty0 054004, 2015.
\newblock \doi{10.1103/PhysRevD.91.054004}.

\bibitem[Nishimura et~al.(2014)Nishimura, Ogilvie, and
  Pangeni]{Nishimura:2014rxa}
Hiromichi Nishimura, Michael~C. Ogilvie, and Kamal Pangeni.
\newblock {Complex saddle points in QCD at finite temperature and density}.
\newblock \emph{Phys. Rev.}, D90\penalty0 (4):\penalty0 045039, 2014.
\newblock \doi{10.1103/PhysRevD.90.045039}.

\bibitem[Akerlund et~al.(2016)Akerlund, de~Forcrand, and
  Rindlisbacher]{Akerlund:2016myr}
Oscar Akerlund, Philippe de~Forcrand, and Tobias Rindlisbacher.
\newblock {Oscillating propagators in heavy-dense QCD}.
\newblock \emph{JHEP}, 10:\penalty0 055, 2016.
\newblock \doi{10.1007/JHEP10(2016)055}.

\bibitem[Bender et~al.(2017)Bender, Felski, Hassanpour, Klevansky, and
  Beygi]{Bender:2016vdo}
Carl~M. Bender, Alexander Felski, Nima Hassanpour, S.~P. Klevansky, and Alireza
  Beygi.
\newblock {Analytic structure of eigenvalues of coupled quantum systems}.
\newblock \emph{Phys. Scripta}, 92\penalty0 (1):\penalty0 015201, 2017.
\newblock \doi{10.1088/0031-8949/92/1/015201}.

\bibitem[Gross and Neveu(1974)]{Gross:1974jv}
David~J. Gross and Andre Neveu.
\newblock {Dynamical Symmetry Breaking in Asymptotically Free Field Theories}.
\newblock \emph{Phys. Rev.}, D10:\penalty0 3235, 1974.
\newblock \doi{10.1103/PhysRevD.10.3235}.

\bibitem[Seul and Andelman(1995)]{Seul476}
Michael Seul and David Andelman.
\newblock Domain shapes and patterns: The phenomenology of modulated phases.
\newblock \emph{Science}, 267\penalty0 (5197):\penalty0 476--483, 1995.
\newblock ISSN 0036-8075.
\newblock \doi{10.1126/science.267.5197.476}.
\newblock URL \url{http://science.sciencemag.org/content/267/5197/476}.

\bibitem[Nussinov et~al.(1999)Nussinov, Rudnick, Kivelson, and
  Chayes]{Nussinov:1999fu}
Zohar Nussinov, Joseph Rudnick, Steven~A. Kivelson, and L.~N. Chayes.
\newblock Avoided critical behavior in $\mathit{O}(\mathit{n})$ systems.
\newblock \emph{Phys. Rev. Lett.}, 83:\penalty0 472--475, Jul 1999.
\newblock \doi{10.1103/PhysRevLett.83.472}.
\newblock URL \url{https://link.aps.org/doi/10.1103/PhysRevLett.83.472}.

\bibitem[Muratov(2002)]{PhysRevE.66.066108}
C.~B. Muratov.
\newblock Theory of domain patterns in systems with long-range interactions of
  coulomb type.
\newblock \emph{Phys. Rev. E}, 66:\penalty0 066108, Dec 2002.
\newblock \doi{10.1103/PhysRevE.66.066108}.
\newblock URL \url{https://link.aps.org/doi/10.1103/PhysRevE.66.066108}.

\bibitem[{Ortix} et~al.(2008){Ortix}, {Lorenzana}, and {di
  Castro}]{PhysRevLett.100.246402}
C.~{Ortix}, J.~{Lorenzana}, and C.~{di Castro}.
\newblock {Coulomb-Frustrated Phase Separation Phase Diagram in Systems with
  Short-Range Negative Compressibility}.
\newblock \emph{Physical Review Letters}, 100\penalty0 (24):\penalty0 246402,
  June 2008.
\newblock \doi{10.1103/PhysRevLett.100.246402}.

\end{thebibliography}

\end{document}